\documentclass [10pt]{article}
\usepackage{graphicx}
\usepackage{amsmath, amsthm, amssymb}

\begin{document}
\title{Kochen-Specker Sets with Thirty Rank-Two Projectors in Three-Qubit System}
\author{S.P. Toh  {\footnote{ Email address:
SingPoh.Toh@nottingham.edu.my; singpoh@gmail.com \newline
\hangindent=0.5cm Tel:
+6(03)8924 8628 Fax: +6(03)8924 8017}} \\
\emph{Faculty of Engineering, The University of Nottingham Malaysia Campus}\\
\emph{Jalan Broga, 43500 Semenyih, Selangor Darul Ehsan, Malaysia}}

\date{}

\maketitle

\begin{center}
\line(1,0){300}
\end{center}

\begin{abstract}

\emph{A simple three rules supplemented by five steps scheme is proposed to produce Kochen-Specker (KS) sets with 30 rank-2 projectors that occur twice each. The KS sets provide state-independent proof of KS theorem based on a system of three qubits. A small adjustment of the scheme enables us to manually generate a large
number of KS sets with a mixture of rank-1 and rank-2 projectors.}

\end{abstract}

\emph{Keywords}: Kochen-Specker theorem; Contextuality; Hidden
variable; Three-qubit.

\begin{center}
\line(1,0){300}
\end{center}

\section{Introduction} \label{Section1}
The Kochen-Specker (KS) theorem demonstrates the inconsistency between predictions
of quantum mechanics (QM) and noncontextual hidden-variable (NCHV) theories. Contextuality is one of the classically unattainable
features of QM. The results of measurements in QM depend on context and do not reveal
preexisting values. A context is a set of maximally collection of compatible observables. The results of measurements
in QM depend on the choice of other compatible measurements that are carried out previously or simultaneously.
The simplest system that can be used to prove KS theorem is a single qutrit. As a qutrit does not refer to nonlocality,
it shows that KS theorem is a more general theorem compare to Bell theorem that rules out the local hidden variable model of QM.

\vspace{2mm}

The possibility of testing KS theorem experimentally was once doubted due to the finiteness in measurement times and
precision \cite{R1, R2}. Cabello \cite{R3} and others \cite{R4} suggested how KS theorem might be experimentally tested by deriving
a set of noncontextual inequalities that are violated by QM for any quantum states but are satisfied by any NCHV theories. Recently,
there are many successful experiments that show the violation of noncontextual inequality, for example the experiments on a pair
of trapped ions \cite{R5}, neutrons \cite{R6}, single photons \cite{R7}, two photonic qubits \cite{R8} and nuclear spins \cite{R9}.

\vspace{2mm}

The original proof of KS theorem involves 117 directions in three-dimensional real Hilbert space \cite{R10}. Peres \cite{R11}
found a simpler proof with 33 and 24 rays for three- and four-dimensional systems, respectively. Mermin \cite{R12} used an
array of nine observables for two spin-$\frac{1}{2}$ particles to show quantum contextuality. Similar mathematical simplicity
is also shown in KS theorem proof for the three-qubit eight-dimensional system using ten observables \cite{R12}. Up to now
the smallest numbers of rays required in the proof of KS theorem are 31 \cite{R13}, 18 \cite{R14}
and 36 \cite{R15} in three-, four- and eight-dimensional systems, respectively.

\vspace{2mm}

The KS sets used to prove the KS theorem are difficult to obtain previously. For example,
there is only one KS set reported in \cite{R16} and \cite{R15} with 20 and 36 rays in four- and eight-dimensional real
Hilbert spaces, respectively. Recently, with the aid of computer, the number of KS sets available increases tremendously. For instance,
the number of KS sets with 36 rays in three-qubit system is 320 according to \cite{R17}. In this Letter,
we adopt a set of simple rules supplemented by a few steps to construct KS sets that consist of 30 rank-2 projectors without relying on computer
computation. In Sec.\ \ref{Sec2} a brief introduction to the 25 bases formed by 40 rays of Kernaghan and Peres is given \cite{R15}.
An example is given in Sec.\ \ref{Sec3} to explicitly show the steps to obtain KS sets involving 30 rank-2 projectors from
KS sets formed by 40 rank-1 projectors provided in \cite{R17}. We generalize the steps in Sec.\ \ref{Sec4}
and conclude in Sec.\ \ref{Sec5}.

\section{Kochen-Specker sets with 15 bases formed by 40 rays} \label{Sec2}
For the sake of completeness, we furnish in this section some necessary basic facts prior to a detail discussion on the procedure of constructing
rank-2 projectors (or plane) KS sets.

\vspace{2mm}

Based on the Mermin pentagram that consists of five sets of four mutually commuting operators, Kernaghan and Peres \cite{R15} derived 40 rank-1 projectors (or rays) to form 25 bases, where each of the bases is a set of mutually orthogonal projectors that spans an eight-dimensional real Hilbert space. Table \ref{T1} lists the 40 rank-1 projectors, $R_i$ with \mbox{$i=$1, 2, 3, \textellipsis, 40}, and Table \ref{T2} which is taken from \cite{R17} lists the 25 bases. The first five bases in Table \ref{T2} are called pure bases ($PB_i$, \mbox{$i=$1, 2, \textellipsis, 5}) \cite{R17} and their mixture give rise to remaining hybrid bases ($HB_i$, \mbox{$i=$6, 7, 8, \textellipsis, 25}). Each of the rank-1 projectors occurs once in $PB$ and four times in $HB$.

\vspace{2mm}

\begin{table}[!h]
\caption{The 40 rays derived by  Kernaghan and Peres for KS proof in three-qubit system.
The symbol $\bar{1}$ is used to denote $-1$. \label{T1}}
\begin{center}
\begin{tabular}{|c|c|c|c|c|c|c|c|c|c|}
  \hline
  1 & 10000000 & 9 & 11110000 & 17 & 11001100 & 25 & 10101010 & 33 & 100101$\bar{1}$ 0 \\
  2 & 01000000 & 10 & 11$\bar{1}$$\bar{1}$0000 & 18 & 1100$\bar{1}$$\bar{1}$00 & 26 & 1010$\bar{1}$0$\bar{1}$0 & 34 & 100$\bar{1}$0110 \\
  3 & 00100000 & 11 & 1$\bar{1}$$1\bar{1}$0000 & 19 & 1$\bar{1}$001$\bar{1}$00 & 27 & 10$\bar{1}$010$\bar{1}$0 & 35 & 10010$\bar{1}$10 \\
  4 & 00010000 & 12 & 1$\bar{1}$$\bar{1}$10000 & 20 & 1$\bar{1}$00$\bar{1}$100 & 28 & 10$\bar{1}$0$\bar{1}$010 & 36 & 100$\bar{1}$0$\bar{1}$$\bar{1}$0 \\
  5 & 00001000 & 13 & 00001111 & 21 & 00110011 & 29 & 01010101 & 37 & 0110$\bar{1}$001 \\
  6 & 00000100 & 14 & 000011$\bar{1}$$\bar{1}$ & 22 & 001100$\bar{1}$$\bar{1}$ & 30 & 01010$\bar{1}$$0\bar{1}$ & 38 & 01$\bar{1}$01001 \\
  7 & 00000010 & 15 & 00001$\bar{1}$1$\bar{1}$ & 23 & 001$\bar{1}$001$\bar{1}$ & 31 & 010$\bar{1}$010$\bar{1}$ & 39 & 0$\bar{1}$101001 \\
  8 & 00000001 & 16 & 00001$\bar{1}$$\bar{1}$1 & 24 & 001$\bar{1}$00$\bar{1}$1 & 32 & 010$\bar{1}$0$\bar{1}$01 & 40 & 0$\bar{1}$$\bar{1}$0$\bar{1}$001 \\
  \hline
\end{tabular}
\end{center}
\end{table}

\vspace{2mm}

\begin{table}[!h]
\caption{Bases formed by eight-dimensional rays listed in Table \ref{T1}.  \label{T2}}
\begin{center}
\begin{tabular}{|c|cccccccc|}
  \hline
  \multicolumn{1}{|c|}{Index}& \multicolumn{8}{|c|}{Rays in Basis} \\
  \hline
  1 & 1 & 2 & 3 & 4 & 5 & 6 & 7 & 8 \\
  2 & 9 & 10 & 11 & 12 & 13 & 14 & 15 & 16 \\
  3 & 17 & 18 & 19 & 20 & 21 & 22 & 23 & 24 \\
  4 & 25 & 26 & 27 & 28 & 29 & 30 & 31 & 32 \\
  5 & 33 & 34 & 35 & 36 & 37 & 38 & 39 & 40 \\ \hline
  6 & 1 & 2 & 3 & 4 & 13 & 14 & 15 & 16 \\
  7 & 1 & 2 & 5 & 6 & 21 & 22 & 23 & 24 \\
  8 & 1 & 3 & 5 & 7 & 29 & 30 & 31 & 32 \\
  9 & 1 & 4 & 6 & 7 & 37 & 38 & 39 & 40 \\
  10 & 2 & 3 & 5 & 8 & 33 & 34 & 35 & 36 \\
  11 & 2 & 4 & 6 & 8 & 25 & 26 & 27 & 28 \\
  12 & 3 & 4 & 7 & 8 & 17 & 18 & 19 & 20 \\
  13 & 5 & 6 & 7 & 8 & 9 & 10 & 11 & 12 \\
  14 & 9 & 10 & 13 & 14 & 19 & 20 & 23 & 24 \\
  15 & 9 & 11 & 13 & 15 & 27 & 28 & 31 & 32 \\
  16 & 9 & 12 & 14 & 15 & 34 & 36 & 38 & 39 \\
  17 & 10 & 11 & 13 & 16 & 33 & 35 & 37 & 40 \\
  18 & 10 & 12 & 14 & 16 & 25 & 26 & 29 & 30 \\
  19 & 11 & 12 & 15 & 16 & 17 & 18 & 21 & 22 \\
  20 & 17 & 19 & 21 & 23 & 26 & 28 & 30 & 32 \\
  21 & 17 & 20 & 22 & 23 & 35 & 36 & 37 & 39 \\
  22 & 18 & 19 & 21 & 24 & 33 & 34 & 38 & 40 \\
  23 & 18 & 20 & 22 & 24 & 25 & 27 & 29 & 31 \\
  24 & 25 & 28 & 30 & 31 & 33 & 36 & 37 & 38 \\
  25 & 26 & 27 & 29 & 32 & 34 & 35 & 39 & 40 \\
  \hline
\end{tabular}
\end{center}
\end{table}

\vspace{2mm}

As a result of computer search, Waegell and Aravind \cite{R17} found 64 KS sets that are composed of 40 rays and 15 bases. A manual construction of these 64 KS sets can be found in \cite{R18}. Since these KS sets have 20 rays that occur twice each, 20 rays that occur four times each among its 15 bases, and each base contains 8 rays,  they are labeled as $20_220_4 \textendash 15_8$ \cite{R17}. The 15 bases are contributed by 5 $PB$s and 10 $HB$s. An example of  $20_220_4 \textendash 15_8$ KS sets is given in Table \ref{T3}.

\vspace{2mm}

\begin{table}[!h]
\caption{KS set that consists of 40 rays and 15 bases. The 20 rays that occur four times each are typed in italic and the 20 rays that occur twice each are in plain type.\label{T3}}
\begin{center}
\begin{tabular}{|c|cccccccc|}
  \hline
  \multicolumn{1}{|c|}{Index}& \multicolumn{8}{|c|}{Rays in Basis} \\
  \hline
  1 & \emph{1} & \emph{2} & \emph{3} & 4 & \emph{5} & 6 & 7 & 8 \\
  2 & \emph{9} & 10 & 11 & 12 & \emph{13} & \emph{14} & \emph{15} & 16 \\
  3 & 17 & 18 & \emph{19} & 20 & \emph{21} & 22 & \emph{23} & \emph{24} \\
  4 & 25 & 26 & 27 & \emph{28} & 29 & \emph{30} & \emph{31} & \emph{32} \\
  5 & \emph{33} & \emph{34} & 35 & \emph{36} & 37 & \emph{38} & 39 & 40 \\ \hline
  6 & \emph{1} & \emph{2} & \emph{3} & 4 & \emph{13} & \emph{14} & \emph{15} & 16 \\
  7 & \emph{1} & \emph{2} & \emph{5} & 6 & \emph{21} & 22 & \emph{23} & \emph{24} \\
  8 & \emph{1} & \emph{3} & \emph{5} & 7 & 29 & \emph{30} & \emph{31} & \emph{32} \\
  10 & \emph{2} & \emph{3} & \emph{5} & 8 & \emph{33} & \emph{34} & 35 & \emph{36} \\
  14 & \emph{9} & 10 & \emph{13} & \emph{14} & \emph{19} & 20 & \emph{23} & \emph{24} \\
  15 & \emph{9} & 11 & \emph{13} & \emph{15} & 27 & \emph{28} & \emph{31} & \emph{32} \\
  16 & \emph{9} & 12 & \emph{14} & \emph{15} & \emph{34} & \emph{36} & \emph{38} & 39 \\
  20 & 17 & \emph{19} & \emph{21} & \emph{23} & 26 & \emph{28} & \emph{30} & \emph{32} \\
  22 & 18 & \emph{19} & \emph{21} & \emph{24} & \emph{33} & \emph{34} & \emph{38} & 40 \\
  24 & 25 & \emph{28} & \emph{30} & \emph{31} & \emph{33} & \emph{36} & 37 & \emph{38} \\
  \hline
\end{tabular}
\end{center}
\end{table}

\vspace{2mm}

The KS sets in the form of $20_220_4 \textendash 15_8$ is constructed completely by rank-1 projectors. However, they can easily be transformed to KS sets that composed merely of rank-2 projectors, see Section \ref{Sec3}.

\section{A Concrete Example: Steps of Construction} \label{Sec3}
Example given in Table \ref{T3} is a KS set that involves 40 rank-1 projectors. We propose in this section steps to transform it to a KS set that involves 30 rank-2 projectors, where each of the projectors occurs twice among the 15 bases, as is shown in Table \ref{T4}.

\vspace{2mm}

\begin{table}[!h]
\caption{KS set consists of 30 rank-2 projectors obtained from the KS set given in Table \ref{T3}.} \label{T4}
\begin{center}
\begin{tabular}{|c|c|c|c|c|}
  \hline
  1 & (\emph{1}, 7) & (\emph{2}, 8) & (\emph{3}, 4) & (\emph{5}, 6) \\ \hline
  2 & (\emph{9}, 12) & (\emph{13}, 16) & (\emph{14}, 10) & (\emph{15}, 11) \\ \hline
  3 & (\emph{19}, 20) & (\emph{21}, 22) & (\emph{23}, 17) & (\emph{24}, 18) \\ \hline
  4 & (\emph{28}, 27) & (\emph{30}, 29) & (\emph{31}, 25) & (\emph{32}, 26) \\ \hline
  5 & (\emph{33}, 35) & (\emph{34}, 40) & (\emph{36}, 37) & (\emph{38}, 39) \\ \hline
  6 & (\emph{1}, \emph{2}) & (\emph{3}, 4) & (\emph{13}, 16) & (\emph{14}, \emph{15}) \\ \hline
  7 & (\emph{1}, \emph{2}) & (\emph{5}, 6) & (\emph{21}, 22) & (\emph{23}, \emph{24}) \\ \hline
  8 & (\emph{3}, \emph{5}) & (\emph{1}, 7) & (\emph{30}, 29) & (\emph{31}, \emph{32}) \\ \hline
  10 & (\emph{3}, \emph{5}) & (\emph{2}, 8) & (\emph{33}, 35) & (\emph{34}, \emph{36}) \\ \hline
  14 & (\emph{14}, 10) & (\emph{9}, \emph{13}) & (\emph{19}, 20) & (\emph{23}, \emph{24}) \\ \hline
  15 & (\emph{15}, 11) & (\emph{9}, \emph{13}) & (\emph{28}, 27) & (\emph{31}, \emph{32}) \\ \hline
  16 & (\emph{9}, 12) & (\emph{14}, \emph{15}) & (\emph{38}, 39) & (\emph{34}, \emph{36}) \\ \hline
  20 & (\emph{23}, 17) & (\emph{19}, \emph{21}) & (\emph{32}, 26) & (\emph{28}, \emph{30}) \\ \hline
  22 & (\emph{24}, 18) & (\emph{19}, \emph{21}) & (\emph{34}, 40) & (\emph{33}, \emph{38}) \\ \hline
  24 & (\emph{31}, 25) & (\emph{28}, \emph{30}) & (\emph{36}, 37) & (\emph{33}, \emph{38}) \\
  \hline
\end{tabular}
\end{center}
\end{table}

\vspace{2mm}

The rank-1 projectors in italic for a specific $PB_i$ form the set $\Gamma^i$, and the remaining rank-1 projectors form the set $\neg \Gamma^i$. Our steps of construction are guided by the following three rules:

\vspace{2mm}

\begin{enumerate}
\item[] Rule 1 ($\Re1$): \\
For $\Gamma^i= \{ \alpha, \beta, \gamma, \delta \}$, we can extract 4 $HB$s that contain subsets labeled by $\Gamma^i_j$, i.e., $\Gamma^i_1=\{ \alpha, \beta, \gamma \}$, $\Gamma^i_2=\{ \alpha, \beta, \delta \}$, $\Gamma^i_3=\{ \alpha, \gamma, \delta \}$ and $\Gamma^i_4=\{ \beta, \gamma, \delta \}$.

\item[] Rule 2 ($\Re2$): \\
Rank-1 projectors from $\Gamma^i$ must be coupled with rank-1 projectors from $\neg \Gamma^i$ to form 4 rank-2 projectors in $PB$ and each of these rank-2 projectors repeats itself once in $HB$.

\item[] Rule 3 ($\Re3$): \\
Rays from $\Gamma^i$ must form 2 rank-2 projectors in $HB$.
\end{enumerate}

\vspace{2mm}

Note that the sequence of the above rules must be taken care of. It is important to apply the rules in the given order,  i.e., $\Re1$ first, followed by $\Re2$ and lastly $\Re3$. Now, let us apply them to our example.

\vspace{2mm}

Step 1 ($S1$) : Take $\Gamma^1 = \{R1, R2, R3, R5\}$. Apply $\Re1$, $\Re2$ and $\Re3$.

The results obtained after the execution of $S1$ are shown in Table \ref{T5}. Note that $\alpha = R1$, $\beta = R2$, $\gamma = R3$ and $\delta = R5$. By applying $\Re1$, we obtained bases 6, 7, 8 and 10. These bases contain $\Gamma^1_1 = \{R1, R2, R3\}$, $\Gamma^2_2 = \{R1, R2, R5\}$, $\Gamma^2_3 = \{R1, R3, R5\}$ and $\Gamma^2_4 = \{R2, R3, R5\}$, respectively.  By applying $\Re2$, namely coupling the rays from $\Gamma^1$ to the rays from $\neg \Gamma^1=\{ R4, R6, R7, R8 \} $, we obtain 4 rank-2 projectors in base 1. Note that the 4 rank-2 projectors in base 1 repeat themselves in the other 4 bases, as shown in Table \ref{T5}. By applying $\Re3$, we obtain rank-2 projectors (\emph{1}, \emph{2}) and (\emph{3}, \emph{5}). All the rank-2 projectors formed are written in parentheses.

\vspace{2mm}

\begin{table}[!h]
\caption{Rank-2 projectors obtained after the execution of $S1$. \label{T5}}
\begin{center}
\begin{tabular}{|c|c|c|c|c|}
  \hline
  1 & (\emph{1}, 7) & (\emph{2}, 8) & (\emph{3}, 4) & (\emph{5}, 6) \\ \hline
  6 & (\emph{1}, \emph{2}) & (\emph{3}, 4) &  &  \\ \hline
  7 & (\emph{1}, \emph{2}) & (\emph{5}, 6) &  &  \\ \hline
  8 & (\emph{3}, \emph{5}) & (\emph{1}, 7) &  &  \\ \hline
  10 &(\emph{3}, \emph{5}) & (\emph{2}, 8) &  &  \\
  \hline
\end{tabular}
\end{center}
\end{table}

\vspace{2mm}

Step 2 ($S2$) : Take $\Gamma^2 = \{R9, R13, R14, R15\}$. Apply $\Re1$, $\Re2$ and $\Re3$.

The results obtained after the execution of $S2$ are shown in Table \ref{T6}. Applying $\Re1$ produces
bases 6, 14, 15 and 16. Applying $\Re2$ produces (\emph{9}, 12), (\emph{13}, 16), (\emph{14}, 10) and
(\emph{15}, 11). Applying $\Re3$ produces (\emph{9}, \emph{13}) and (\emph{14}, \emph{15}). As for the
results of $S1$, carrying out the three rules in $S2$ produces six pairs of rank-2 projectors. Note that (\emph{1}, \emph{2})
and (\emph{3}, 4) in base 6 have been produced prior to the execution of $S2$.

\vspace{2mm}

\begin{table}[!h]
\caption{Rank-2 projectors obtained after the execution of $S2$. \label{T6}}
\begin{center}
\begin{tabular}{|c|c|c|c|c|}
  \hline
  2 & (\emph{9}, 12) & (\emph{13}, 16) & (\emph{14}, 10) & (\emph{15}, 11) \\ \hline
  6 & (\emph{1}, \emph{2}) & (\emph{3}, 4) & (\emph{13}, 16) & (\emph{14}, \emph{15}) \\ \hline
  14 & (\emph{14}, 10) & (\emph{9}, \emph{13}) &  &  \\ \hline
  15 & (\emph{15}, 11) & (\emph{9}, \emph{13}) &  &  \\ \hline
  16 &(\emph{9}, 12) & (\emph{14}, \emph{15}) &  &  \\
  \hline
\end{tabular}
\end{center}
\end{table}

\vspace{2mm}

Step 3 ($S3$) : Take $\Gamma^3 = \{R19, R21, R23, R24\}$. Apply $\Re1$, $\Re2$ and $\Re3$.

The results obtained after the execution of $S3$ are shown in Table \ref{T7}.
Applying $\Re1$ produces bases 7, 14, 20 and 22. Applying $\Re2$ produces (\emph{19}, 20),
(\emph{21}, 22), (\emph{23}, 17) and (\emph{24}, 18). Applying $\Re3$ produces (\emph{19}, \emph{21})
and (\emph{23}, \emph{24}). Note that (\emph{1}, \emph{2}) and (\emph{5}, 6) in base 7 and
(\emph{14}, 10) and (\emph{9}, \emph{13}) in base 14 have been produced prior to the execution of $S3$.

\vspace{2mm}

\begin{table}[!h]
\caption{Rank-2 projectors obtained after the execution of $S3$. \label{T7}}
\begin{center}
\begin{tabular}{|c|c|c|c|c|}
  \hline
  3 & (\emph{19}, 20) & (\emph{21}, 22) & (\emph{23}, 17) & (\emph{24}, 18) \\ \hline
  7 & (\emph{1}, \emph{2}) & (\emph{5}, 6) & (\emph{21}, 22) & (\emph{23}, \emph{24}) \\ \hline
  14 & (\emph{14}, 10) & (\emph{9}, \emph{13}) & (\emph{19}, 20) & (\emph{23}, \emph{24}) \\ \hline
  20 & (\emph{23}, 17) & (\emph{19}, \emph{21}) &  &  \\ \hline
  22 & (\emph{24}, 18) & (\emph{19}, \emph{21}) &  &  \\
  \hline
\end{tabular}
\end{center}
\end{table}

\vspace{2mm}

Step 4 ($S4$) : Take $\Gamma^4 = \{R28, R30, R31, R32\}$. Apply $\Re1$, $\Re2$ and $\Re3$.

The results obtained after the execution of $S4$ are shown in Table \ref{T8}.
Applying $\Re1$ produces bases 8, 15, 20 and 24. Applying $\Re2$ produces (\emph{28}, 27),
(\emph{30}, 29), (\emph{31}, 25) and (\emph{32}, 26). Applying $\Re3$ produces (\emph{28}, \emph{30})
and (\emph{31}, \emph{32}). Note that (\emph{3}, \emph{5}) and (\emph{1}, 7) in base 8,
(\emph{15}, 11) and (\emph{9}, \emph{13}) in base 15 and (\emph{23}, 17) and (\emph{19}, \emph{21})
in base 20 have been produced prior to the execution of $S4$.

\vspace{2mm}

\begin{table}[!h]
\caption{Rank-2 projectors obtained after the execution of $S4$. \label{T8}}
\begin{center}
\begin{tabular}{|c|c|c|c|c|}
  \hline
  4 & (\emph{28}, 27) & (\emph{30}, 29) & (\emph{31}, 25) & (\emph{32}, 26) \\ \hline
  8 & (\emph{3}, \emph{5}) & (\emph{1}, 7) & (\emph{30}, 29) & (\emph{31}, \emph{32}) \\ \hline
  15 & (\emph{15}, 11) & (\emph{9}, \emph{13}) & (\emph{28}, 27) & (\emph{31}, \emph{32}) \\ \hline
  20 & (\emph{23}, 17) & (\emph{19}, \emph{21}) & (\emph{32}, 26) & (\emph{28}, \emph{30}) \\ \hline
  24 & (\emph{31}, 25) & (\emph{28}, \emph{30}) &  &  \\
  \hline
\end{tabular}
\end{center}
\end{table}

\vspace{2mm}

Step 5 ($S5$) : Take $\Gamma^5 = \{R33, R34, R36, R38\}$. Apply $\Re1$, $\Re2$ and $\Re3$.

The results obtained after the execution of $S5$ are shown in Table \ref{T9}.
Applying $\Re1$ produces bases 10, 16, 22 and 24. Applying $\Re2$ produces (\emph{33}, 35),
(\emph{34}, 40), (\emph{36}, 37) and (\emph{38}, 39). Applying $\Re3$ produces (\emph{33}, \emph{38})
and (\emph{34}, \emph{36}). Note that (\emph{3}, \emph{5}) and (\emph{2}, 8) in base 10,
(\emph{9}, 12) and (\emph{14}, \emph{15}) in base 16, (\emph{24}, 18) and (\emph{19}, \emph{21})
in base 22 and (\emph{31}, 25) and (\emph{28}, \emph{30}) in base 24 have been produced
prior to the execution of $S5$.

\vspace{2mm}

\begin{table}[!h]
\caption{Rank-2 projectors obtained after the execution of S5. \label{T9}}
\begin{center}
\begin{tabular}{|c|c|c|c|c|}
  \hline
  5 & (\emph{33}, 35) & (\emph{34}, 40) & (\emph{36}, 37) & (\emph{38}, 39) \\ \hline
  10 & (\emph{3}, \emph{5}) & (\emph{2}, 8) & (\emph{33}, 35) & (\emph{34}, \emph{36}) \\ \hline
  16 & (\emph{9}, 12) & (\emph{14}, \emph{15}) & (\emph{38}, 39) & (\emph{34}, \emph{36}) \\ \hline
  22 & (\emph{24}, 18) & (\emph{19}, \emph{21}) & (\emph{34}, 40) & (\emph{33}, \emph{38}) \\ \hline
  24 & (\emph{31}, 25) & (\emph{28}, \emph{30}) & (\emph{36}, 37) & (\emph{33}, \emph{38}) \\
  \hline
\end{tabular}
\end{center}
\end{table}

\vspace{2mm}

Table \ref{T5} to Table \ref{T9} list in parentheses the rank-2 projectors formed after the execution of $S1$ to $S5$, respectively, and it is conspicuous that there are overlapping bases. After the completion of the five steps, we extract every different bases once, and for those that occur more than once, we pick the one that is maximally filled. The result obtained would be a KS set shown in Table \ref{T4}. As there are 30 rank-2 projectors and each of them
occurs twice among the 15 bases, the KS set obtained can be used to provide state independent parity proof of the KS theorem.

\section{Discussion} \label{Sec4}
The scheme proposed in Sec.\ \ref{Sec3} is conceived based on the properties shared by all KS sets in the type of $20_220_4 \textendash 15_8$. Apart from the features reflected by the symbol $20_220_4 \textendash 15_8$, we would like to stress that these 15 bases must be composed of 5 $PB$s an 10 $HB$s. Most importantly, the 20 rays that repeat four times each provide us clues to form the rank-2 projectors. Due to the common features shared, $S1$ to $S5$ used to
construct KS set of 30 rank-2 projectors in Sec.\ \ref{Sec3} can be generalized and apply to all 64 KS sets with $20_220_4 \textendash 15_8$, as follows,

\vspace{2mm}

\indent \indent Step 1 ($S1^\prime$) : Apply $\Re1$, $\Re2$ and $\Re3$ to $\Gamma^1$. \\
\indent \indent Step 2 ($S2^\prime$) : Apply $\Re1$, $\Re2$ and $\Re3$ to $\Gamma^2$. \\
\indent \indent Step 3 ($S3^\prime$) : Apply $\Re1$, $\Re2$ and $\Re3$ to $\Gamma^3$. \\
\indent \indent Step 4 ($S4^\prime$) : Apply $\Re1$, $\Re2$ and $\Re3$ to $\Gamma^4$. \\
\indent \indent Step 5 ($S5^\prime$) : Apply $\Re1$, $\Re2$ and $\Re3$ to $\Gamma^5$. \\

\vspace{2mm}

In $S1$ of the example in Sec. \ref{Sec3}, there are in fact three ways to form rank-2 projectors while applying $\Re2$ to base 1.
Specifically, R1 can couple either with R4, R6 or R7 to form (\emph{1}, 4), (\emph{1}, 6) or (\emph{1}, 7), respectively.
On the other hand, (\emph{1}, 8) is disallowed as it doest not appear the second time in any bases of 6, 7, 8 or 10. The
application of $\Re3$ in $S1$ corresponding to the options of (\emph{1}, 4), (\emph{1}, 6) or (\emph{1}, 7) produces three pair
of  rank-2 projectors, i.e., (\emph{1}, \emph{5}) and (\emph{2}, \emph{3}), (\emph{1}, \emph{3}) and (\emph{2}, \emph{5}) or (\emph{1}, \emph{2})
and (\emph{3}, \emph{5}), respectively. Similar situation happens during $S2$ to $S4$ as well. Therefore, based on the generalization,
we know that there are three ways in each step, from $S1^\prime$ to $S5^\prime$, to form 6 rank-2 projectors and the total number of KS sets
of 30 rank-2 projectors that are transformed from each of the KS sets in the type of $20_220_4 \textendash 15_8$ is $3^5=243$.

\vspace{2mm}

Each application of $\Re2$ and $\Re3$ produces 4 and 2 rank-2 projectors, respectively. This clearly explains why there are
in total 30 rank-2 projectors formed upon the completion of $S1^\prime$ to $S5^\prime$.  However, there are various combinations
of invalidate or removing $\Re2$ or $\Re3$ through out the process of construction in order to obtain various numbers,
ranging from two to thirty, of rank-2 projectors. Let us now consider one of the scenarios and investigate how, without $\Re3$, the
number of rank-2 projectors is affected. The aforementioned scheme needs to be further generalized as follow,

\vspace{2mm}

\indent \indent Step 1 ($S1^{\prime \prime}$) : Apply $\Re1$ and $\Re2$ to $\Gamma^1$. Check if $\Re3$ is applicable. \\
\indent \indent Step 2 ($S1^{\prime \prime}$) : Apply $\Re1$ and $\Re2$ to $\Gamma^2$. Check if $\Re3$ is applicable. \\
\indent \indent Step 3 ($S1^{\prime \prime}$) : Apply $\Re1$ and $\Re2$ to $\Gamma^3$. Check if $\Re3$ is applicable. \\
\indent \indent Step 4 ($S1^{\prime \prime}$) : Apply $\Re1$ and $\Re2$ to $\Gamma^4$. Check if $\Re3$ is applicable. \\
\indent \indent Step 5 ($S1^{\prime \prime}$) : Apply $\Re1$ and $\Re2$ to $\Gamma^5$. Check if $\Re3$ is applicable. \\

\vspace{2mm}

Note that if $\Re3$ is applicable, it increases the number of rank-2 projectors formed by two every time we apply it.

\vspace{2mm}

In the $S2$ of our example (cf.\ Sec.\ \ref{Sec3}), the choice of rank-2 projectors for base 2 shown in Table \ref{T6} guarantees the
applicability of $\Re3$. There are two more ways that make the $\Re3$ applicable in $S2$.
However, we can, for example, choose (\emph{9}, 10), (\emph{13}, 16), (\emph{14}, 12) and (\emph{15}, 11), for base 2 instead,
but it will then make $\Re3$ inapplicable. There are in total six ways of forming rank-2 projectors for base 2 that make $\Re3$ inapplicable.
Table \ref{T10} lists all the nine ways of forming rank-2 projectors for base 2. The same situation happens in $S3$ to  $S5$ as well.

\vspace{2mm}

\begin{table}[!h]
\caption{Each of the nine rows shows different way of forming rank-2 projectors for base 2 as a result of
applying $\Re2$. The first three ways make $\Re3$ applicable while the other six ways render $\Re3$ fails. The first way
shown in the first row is the one adopted in Table \ref{T6}.
\label{T10}}
\begin{center}
\begin{tabular}{|c|c|c|c|}
  \hline
  (\emph{9}, 12) & (\emph{13}, 16) & (\emph{14}, 10) & (\emph{15}, 11) \\ \hline
  (\emph{9}, 11) & (\emph{13}, 10) & (\emph{14}, 16) & (\emph{15}, 12) \\ \hline
  (\emph{9}, 10) & (\emph{13}, 11) & (\emph{14}, 12) & (\emph{15}, 16) \\ \hline
  (\emph{9}, 10) & (\emph{13}, 16) & (\emph{14}, 12) & (\emph{15}, 11) \\ \hline
  (\emph{9}, 11) & (\emph{13}, 16) & (\emph{14}, 10) & (\emph{15}, 12) \\ \hline
  (\emph{9}, 10) & (\emph{13}, 11) & (\emph{14}, 16) & (\emph{15}, 12) \\ \hline
  (\emph{9}, 12) & (\emph{13}, 10) & (\emph{14}, 16) & (\emph{15}, 11) \\ \hline
  (\emph{9}, 11) & (\emph{13}, 10) & (\emph{14}, 12) & (\emph{15}, 16) \\ \hline
  (\emph{9}, 12) & (\emph{13}, 11) & (\emph{14}, 10) & (\emph{15}, 16) \\
  \hline
\end{tabular}
\end{center}
\end{table}

In the scenarios where $\Re2$ and $\Re3$ are both applicable, we always have the freedom to choose not to apply
$\Re3$ after the execution of $\Re2$, depends on how many rank-2 projectors we aim to get in the transformed KS sets.
However, in $S1$, as mentioned before, there are three ways of applying $\Re2$ on base 1 that guarantee the
applicability of $\Re3$ and none of the cases make $\Re2$ satisfied and $\Re3$ dissatisfied.
Again, our analysis of the example in Sec. \ref{Sec3} can be generalized to $S1^{\prime \prime}$ to $S5^{\prime \prime}$.
In short, there are three (six) ways of forming 4 rank-2 projectors in $S1^{\prime \prime}$ (each of $S2^{\prime \prime}$ to $S5^{\prime \prime}$)
by applying $\Re2$ and not to execute $\Re3$ although it is applicable, three ways of forming 6 (4+2) rank-2 projectors in each of
$S1^{\prime \prime}$ to $S5^{\prime \prime}$ by applying both $\Re2$ and $\Re3$ and six ways of forming
4 rank-2 projectors in each of $S2^{\prime \prime}$ to $S5^{\prime \prime}$ by applying only $\Re2$ due to the inapplicability of $\Re3$.
Table \ref{T11} shows the numbers of KS sets with various numbers of rank-1 and rank-2 projectors that can be generated
via the adjustment on the number of times $\Re3$ is applied throughout $S2^{\prime \prime}$ to $S5^{\prime \prime}$ (we always apply $\Re3$ on
$S1^{\prime \prime}$ for the ease of computation in Table \ref{T11}). Note that as $N_{\Re3}$ does not reflect specifically at which step the $\Re3$ is
inapplicable or not to be executed (in the case of $\Re3$ is applicable), the result of $N_{KS}$ shown is for only one case.

\vspace{2mm}

So far we consider only one of the examples of KS sets in the form of $20_220_4 \textendash 15_8$, it is obvious that the number of KS sets with the mixture of rank-1 and rank-2 projectors that can be generated from our scheme is indeed huge. Finally, note that when $N_{\Re3}=0$,
$S1^{\prime \prime}$ to $S5^{\prime \prime}$ reduced to $S1$ to $S5$ , and $N_{KS}=243$ is the
same as the number of KS sets we deduced before in our example.

\vspace{2mm}

\begin{table}[!h]
\caption{The number of KS sets generated by applying $\Re1$ and $\Re2$ while invalidating or not executing $\Re3$ throughout $S2^{\prime \prime}$ to $S5^{\prime \prime}$. Note that $\Re3$ is always executed on $S1^{\prime \prime}$ here. The symbols $N_{\Re3}$, N$_{KS}$, $N_{2}$ and $N_{1}$ denote the number of times $\Re3$ is invalidated or not executed, the number of KS sets generated, the number of rank-2 projectors formed and the number of the remaining rank-1 projectors, respectively.
\label{T11}}
\begin{center}
\begin{tabular}{|c|c|c|c|}
  \hline
  $N_{\Re3}$ & $N_{KS}$ & $N_{2}$ & $N_{1}$ \\ \hline
  0 & $3^5 \times 6^0 = 243$ & 30 & 0 \\ \hline
  1 & $3^4 \times 6 = 486$  & 28 & 4 \\ \hline
  2 & $3^3 \times 6^2 = 972$ & 26 & 8 \\ \hline
  3 & $3^2 \times 6^3 = 1944$ & 24 & 12 \\ \hline
  4 & $3 \times 6^4 = 3888$ & 22 & 16 \\
  \hline
\end{tabular}
\end{center}
\end{table}

\section{Conclusion} \label{Sec5}
We proposed a simple scheme of three rules supplemented by five steps to transform the $20_220_4 \textendash 15_8$ Kochen-Specker (KS) sets
into KS sets that involve a mixture of rank-1 and rank-2 projectors. A concrete example is provided as illustration. By manipulating the rules  throughout the five steps, we can determine the number of rank-2 projectors formed in the resultant KS sets. The simplest result obtained is the KS sets with 30 rank-2 projectors that occur twice each among 15 bases. To our knowledge, this is the first rank-2 projectors KS sets produced for three-qubit system based on the Mermin's pentagram.
It can be cast in the form of testable inequality proposed by Cabello (see first inequality in \cite{R3}) . It is also noteworthy that a
considerable number of KS sets can be generated by our scheme without resorting to any computer calculation.

\section*{Acknowledgements}
 The author thanks B. A. Tay for improving the English in the manuscript. This work is supported by the Ministry of Higher Education of Malaysia (MOHE) under the FRGS grant FRGS/1/2011/ST/UNIM/03/1.

\end{document}